# Photo-thermal Self-oscillations in Cavity-Coupled Carbon Nanotube *pn*-Devices

*Moh. R. Amer*[1,2], Tony Levi[3,5], and Stephen B. Cronin*[3,4,5]*

Department of Electrical Engineering[1]
University of California, Los Angeles, California, 90095
Center of Excellence for Green Nanotechnologies[2]
King Abdulaziz City for Science and Technology, Riyadh, 12612, Saudi Arabia
Departments of and Electrical Engineering[3], Material Science[4], Physics and Astronomy[5],
University of Southern California, Los Angeles, CA, 90089

**Abstract:**

We observe photothermal self-oscillations in individual, suspended, quasi-metallic carbon nanotube (CNT) *pn* devices irradiated with focused CW 633nm light. Here, the bottom of the trench forms an optical cavity with an anti-node at $\lambda/4$. Oscillations arise from the optical heating of the nanotube, which causes thermal contraction of the nanotube (negative thermal expansion coefficient). This, in turn, moves the CNT out of the anti-node (maximum field intensity), where the nanotube cools to a lower temperature. It then expands and returns to the maximum field intensity anti-node where it is optically heated once again. The oscillations are observed through a change of the tunneling current in the CNT device. A *pn*-junction, established by two electrostatic gates positioned beneath the nanotube, results in Zener tunneling, which depends strongly on temperature. A Zener tunneling model with oscillating temperature shows good agreement with our measured *I-V* curves, providing further evidence that these oscillations are photothermal in nature.

Coupling mechanical resonators to optically-resonant cavities can result in several useful phenomena, including self-oscillations, side-band cooling, and quantum zero-point motion.[1, 2] Aluminum membrane MEMS resonators have demonstrated resonance frequencies up to 7.5 GHz.[1] Side-band cooling and zero-point motion have recently been demonstrated in this system at mK temperatures.[1] The small mass density ($\mu$ = 5 ag/μm) and high stiffness (i.e., Young's modulus, $E$ = 1 TPa) of carbon nanotubes and graphene provide a unique system for NEMS resonators, with a strength-to-weight ratio three orders of magnitude higher than most known bulk materials. $Q$-factors exceeding $10^5$ have been demonstrated in these devices [3-5]. It has been shown that a discrete drop in the mechanical resonance frequency of these devices on the order of 5 – 20 MHz, as the temperature is varied.[6] This discrete drop is due to a hysteretic clamping/unclamping of the nanotube from the trench sidewall caused van der Waals bonding. For a 1 μm device, the drop observed in resonance frequency corresponds to a change in nanotube length of approximately 50 - 65 nm. While optically driven graphene resonators have been studied since 2007[2, 7], no previous reports of optically driven CNT resonators has been given.

In our previous work, we demonstrated rectifying behavior of quasi-metallic *pn*-junctions at cryogenic temperatures[8]. This work showed that quasi-metallic nanotubes exhibit tunable breakdown voltages at relatively low bias. The mechanism underlying this breakdown was identified as Zener tunneling, where electrons tunnel across the band gap from the *p*- to the *n*-type regions. The CNT *pn*-junction provides a unique ability to further utilize this opto-mechanical system. Here, we present the first demonstration of this functionality.

In the work presented here, low frequency photo-thermal oscillations are detected by coupling a suspended carbon nanotube to an optical cavity. The oscillations are measured through changes induced in the electric current measured through the *pn*-device. In order to understand the fundamental mechanism underlying these oscillations, a model is developed by coupling Zener tunneling with thermal oscillations.

Figures 1a and 1b show a schematic diagram and an SEM image of the device structure, respectively. As indicated in these figures, two gate electrodes are positioned beneath an individual, suspended quasi-metallic carbon nanotube in order to electrostatically create *p*- and *n*-type regions. Devices are fabricated by first depositing two 1µm wide Pt gate electrodes with 200nm separation. Silicon oxide (500nm thick) and silicon nitride (100nm thick) are then grown on top of the gates and etched to create 1µm trenches that expose the two split gate electrodes. Pt source and drain electrodes are patterned on top of the silicon nitride, and Mo catalyst is deposited on top of the source and drain electrodes to initiate the nanotube growth at 875$^{o}$C for 10 minutes, as described previously[9-12]. The nanotube growth is the final step in this sample fabrication process, which ensures that these nanotubes are not contaminated by any chemical residues from the lithographic fabrication processes. Optoelectronic measurements of these devices were taken in a high vacuum ($10^{-6}$ Torr) cryogenic chamber at 4K.

Figure 2 shows the *I-V* characteristics of a suspended quasi-metallic *pn*-junction gated with $V_{g1} = -V_{g2} = 9$V. The dark *I-V* curve shows rectifying behavior with Zener breakdown at a reverse bias of -1.2V, consistent with our previous work on Zener tunneling[8]. The inset shows the $I$-$V_g$ characteristics taken with $V_{g1}$=$V_{g2}$ showing a large modulation of the conductance, indicating a relatively large band gap for this quasi-metallic nanotube[9]. The red curve in Figures 2a and 2b show the *I-V* characteristics taken under illumination with 6.9mW laser light focused to a 1µm spot. Under laser illumination, clear oscillations can be seen in the *I-V* curve with a period of approximately 30mV. These oscillations were observed while sweeping the bias voltage at a rate of 1.02mV/msec and actually correspond to a time-varying oscillation rather than voltage-dependent oscillation (e.g., Coulomb blockade). Similar data from another suspended CNT *pn*-device is shown in figure S1 of the supplementary document. After considering several schemes including Coulomb blockade and photon-assisted tunneling between *n*- and *p*-quantum dots, we believe these oscillations arise from photothermal self-oscillations, which are observed in the *I*-*V* characteristics simply because we are sweeping the voltage at a constant rate. Figure 3a shows the time dependence of the electric current taken at a fixed bias voltage of -1V. Here, a sinusoidal time varying signal can be seen with a period of oscillation of 628 msec, as determined by taking a Fourier transform, as shown in Figure 3b. Here, a low pass filter is used to eliminate high frequency noise. The measured time scale is consistent with the scanning rate and periods observed in the voltage sweeps of Figure 2.

Figure 4a shows the temperature dependence of the *I*-*V* characteristics measured without illumination. Here, the rectifying behavior subsides for temperatures above 10K. Figure 4b shows the *I*-*V* characteristics taken on the same sample with 6.9 and 17mW of incident laser power. The striking similarity of these two datasets (Figures 4a and 4b) indicate that the origin of these oscillations is indeed photothermal in nature. Furthermore, 6.9mW is an extremely high laser intensity, which is known to cause significant heating in these suspended nanotubes, and burnout under ambient (room temperature) conditions.[13] It is, therefore, inconceivable that substantial heating is not taking place in these devices. The photothermal self-oscillation mechanism proceeds as follows: The bottom of the trench forms an optical cavity with an anti-node (maximum field intensity) at $\lambda/4\approx158$nm. Optical heating causes thermal contraction (negative thermal expansion coefficient), which in turn moves the CNT out of the anti-node. When the nanotube cools to a lower temperature, it expands and returns to the maximum field intensity anti-node where it is optical heated once again, and so on. This mechanism is depicted schematically in Figure 5a.

In order to further understand the underlying mechanism of these oscillations, we created a model that couples the heating effect (temperature) with Zener tunneling. Briefly, the Zener tunneling occurs when electrons in the valence band of the *p*-type region tunnel across the band gap ($E_{\text{gap}}$) to the conduction band of the *n*-type region [14]. The WKB tunneling probability for this process is given by

$$T_{WKB} = \exp\left[\frac{-4E_{\text{gap}}^2}{3\sqrt{2}\hbar\upsilon_F\mathcal{E}}\right], \tag{1}$$

where $\mathcal{E}$ is the electric field in the junction which depends on built-in voltage and the tunneling length, and $\upsilon_F$ is the Fermi velocity (8.4 × 105 m/s). The *I-V* characteristics can be obtained by calculating the total current according to (2)

$$I = \frac{4e}{h}\int_{-\infty}^{\infty} T_{WKB}[F_v(E - eV) - F_c(E)]\, dE,$$

where *F* is the temperature dependent Fermi Dirac distribution and *V* is the applied bias voltage. Figure 5c shows the *I-V* curves for a nanotube that is thermally oscillating between two different temperatures (200K and 350K) calculated using the Zener model. The result of these calculations is in excellent agreement with the experimental data, as shown in Figure 5b. Here, the acute temperature sensitivity of the Zener tunneling enables us to read out these temperature oscillations electronically.

The mechanical resonance frequencies of ~1µm-long suspended carbon nanotubes lie in the range of 10-100MHz[15], and the thermal time constant of these devices has been estimated to be even higher.[6] Even the ring up and ring down times of these resonators ($t$~$Q/f_o$, $Q$=$10^3$, $f_o$=$10^6$) should be on the order of 1kHz. While it is likely that these high frequency resonances exist, our experimental measurements are only sensitive to long period oscillations. There are several possible explanations for the extremely long periods of oscillation that we are observing here. One possibility is that beating between two modes that are close in frequency, perhaps due to a small perturbation along the length of the nanotubes, which breaks the symmetry of the fundamental mode. However, given the finite linewidth of these mechanical resonances ($1/Q$ ~ kHz), it is unlikely that beating could produce a coherent oscillation at 1Hz. Another, more likely, explanation of these slower periods of oscillation involves a hysteretic change in the clamping conditions at the ends of the nanotube. In our previous work, we showed that thermal cycling of these suspended nanotube devices results in a hysteretic clamping/unclamping due to a competition between van der Waals forces and thermal fluctuations in the suspended nanotube.[16] We believe that this hysteretic clamping/unclamping could be the primary mechanism underlying these oscillations, which can certainly occur at these long times scales.

In conclusion, we observe photothermal self-oscillations in individual suspended quasi-metallic carbon nanotubes (CNT) irradiated with intense focused CW light (6.9mW). The bottom of the trench forms an optical cavity with an anti-node (maximum) at $\lambda/4 \approx 158$nm. Self-oscillations ensue from the optical heating of the nanotube, which causes thermal contraction (negative TEC), which in turn moves the CNT out of the anti-node (maximum field intensity). When the nanotube cools to a lower temperature, it expands and returns to the maximum field intensity anti-node where it is optical heated once again, and so on. The oscillations in the nanotube temperature are observed in the *I-V* characteristics of the CNT, which is electrostatically gated in a *pn*-junction configuration. Due to the quasi-metallic CNT's small band gap, this *pn*-junction only shows rectifying behavior at low temperatures near 4K. Upon heating, the *I-V* characteristics evolve from rectifying to "S" shape curves. Here, oscillations in the electric current are observed due to the time varying optical heating, which results in oscillations in the *I-V* characteristics of the CNT, obtained by scanning over a 1 minute bias voltage sweep. Fixing the bias voltage at -1V, we observe a time varying change in the electric current with a period of 628 msec.

## Acknowledgements

The authors would like to thank Dr. John Teufel for valuable discussions. Part of this work performed at USC was supported in part by Department of Energy (DOE) Award No. DE-FG02-07ER46376. A portion of this work was carried out in the University of California Santa Barbara (UCSB) nanofabrication facility, part of the NSF funded National Nanotechnology Infrastructure Network (NNIN). The author would like to acknowledge KACST for their support through the Center of Excellence for Green Nanotechnologies (CEGN).

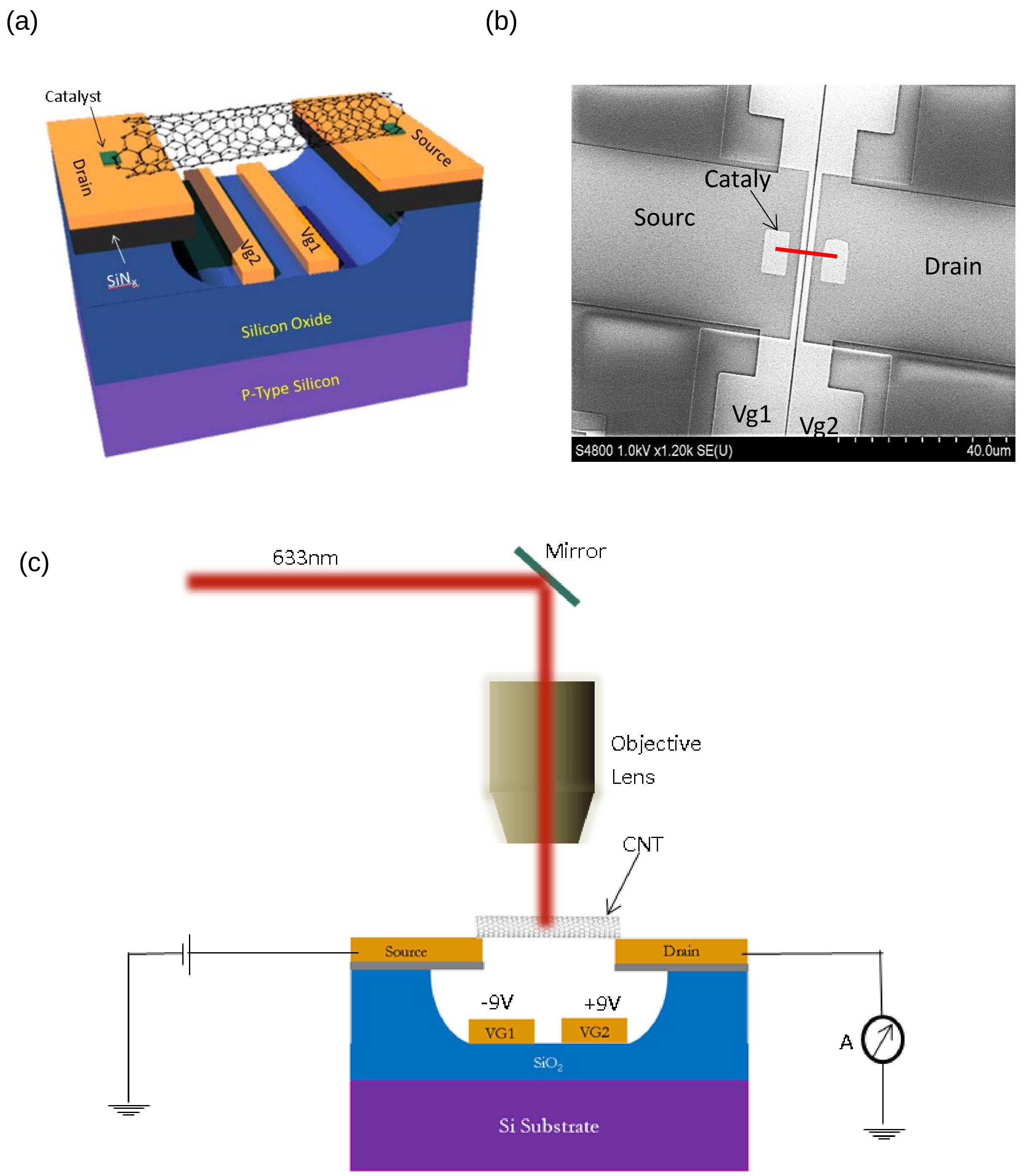

**Figure 1. Device Structure and Experimental Setup**. (a) Schematic diagram showing the metallic carbon nanotube suspended over a trench. (b) SEM image of the device structure. The nanotube location is highlighted in red. (c) Photothermal excitation experimental setup.

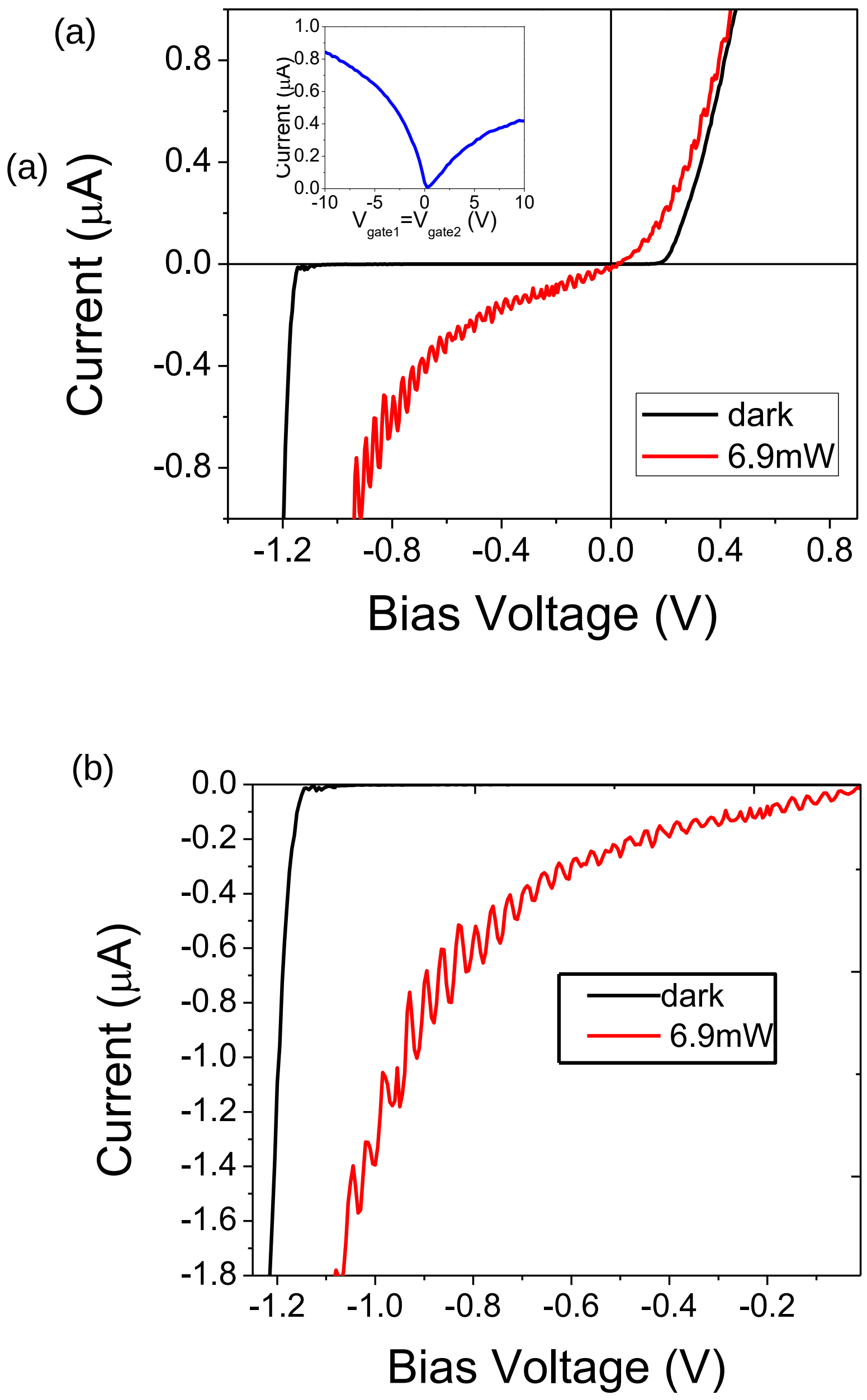

**Figure 2. Current-Voltage Characteristics of Thermal Oscillations.** (a) $I$-$V_{bias}$ characteristics of a quasi-metallic carbon nanotube with and without optical excitation. The inset shows the conductance of the nanotube plotted as a function of gate voltage. (b) A zoomed-in view of the induced current oscillations showing the periodicity of the oscillations.

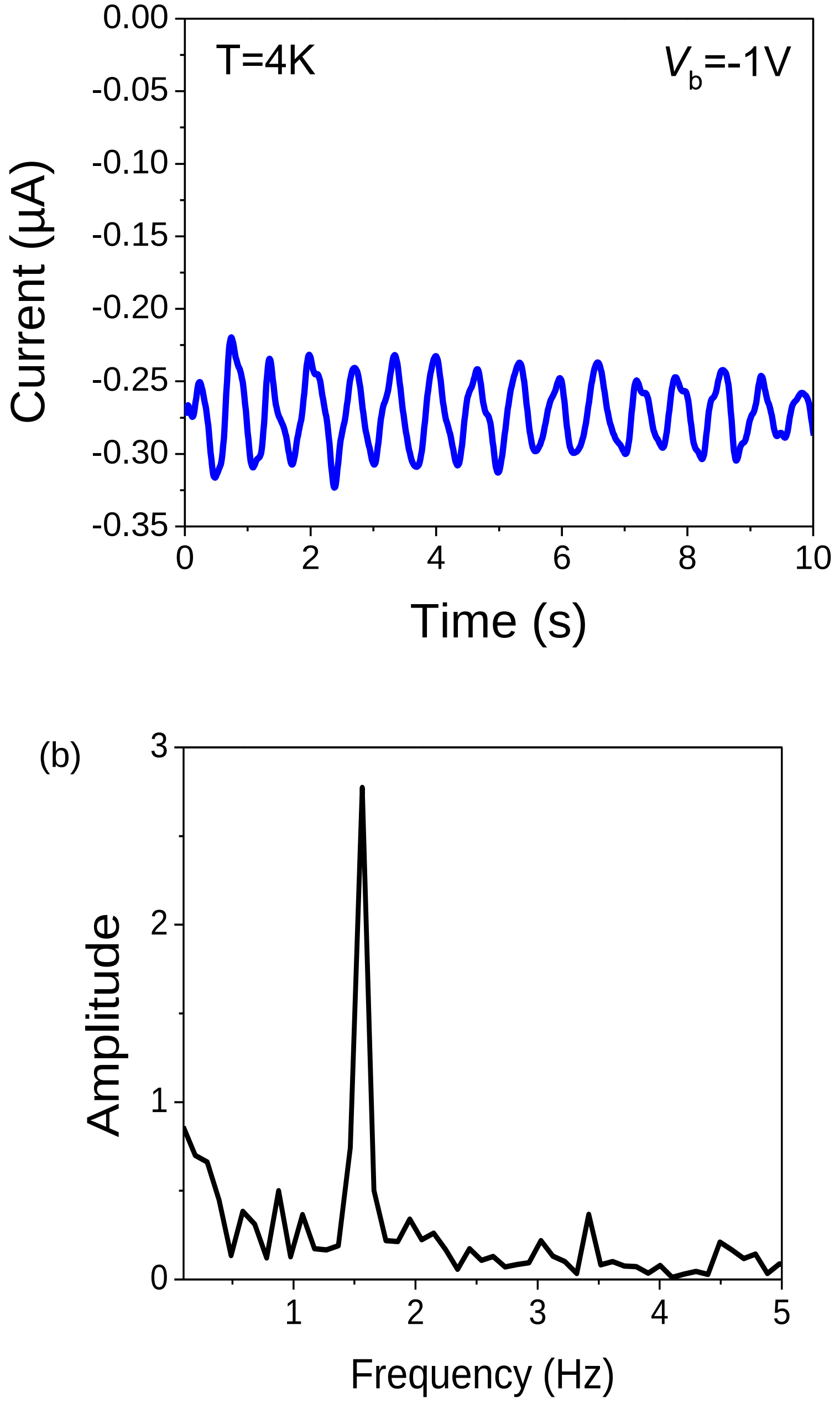

**Figure 3. Low Frequency Measurements of Thermal Oscillations.** (a) Time dependence of the current through the nanotube, corresponding to thermal oscillations as a function of time showing a sinusoidal behavior. (b) Fast Fourier Transform of the data in (a) showing the dominant peak at 1.56 Hz.

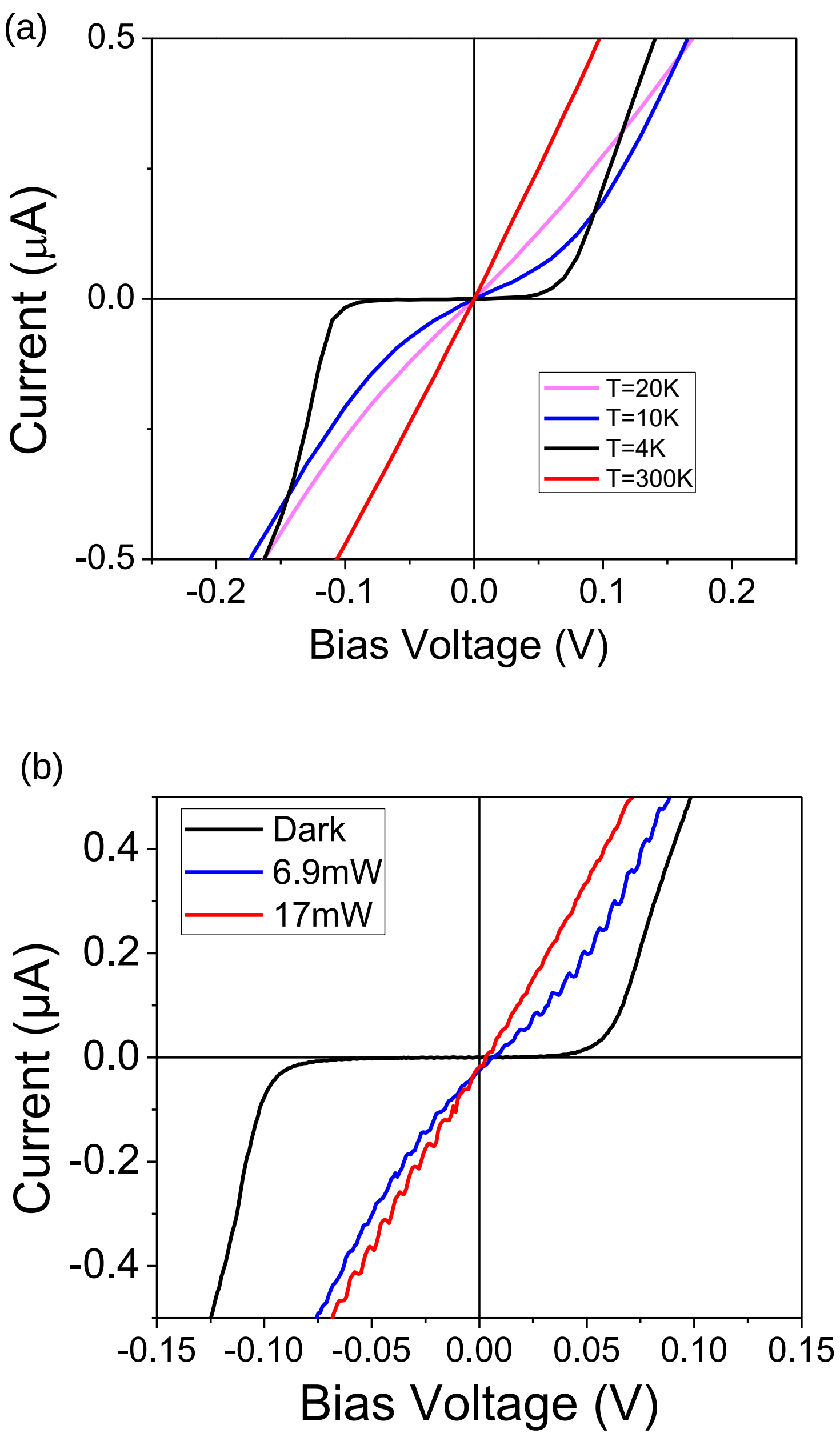

**Figure 4. Laser Power and Temperature Dependence of the *I*-*V* Characteristics.** (a) Temperature dependence of the $I$-$V_{bias}$ curves demonstrating that Zener tunneling is very sensitive to temperature and occurs at T~4K. (b) $I$-$V_{bias}$ characteristics taken in the dark and under optical excitations with a 633nm laser. The measured laser power is 6.9mW (red curve) and 17mW (blue curve).

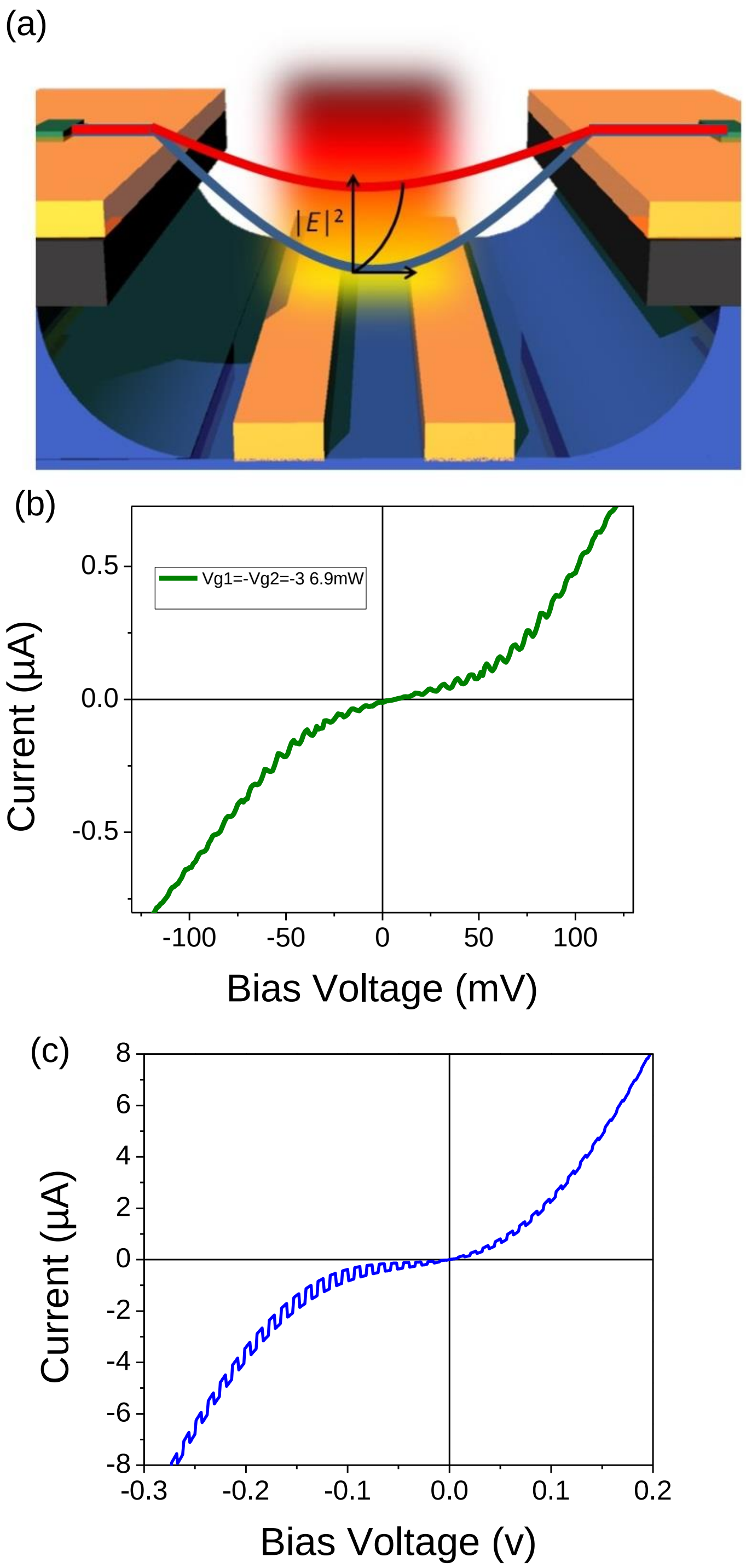

**Figure 5. Coupling of Zener Tunneling Model with the Thermal Oscillations.** (a) Schematic diagram showing photo-thermal self-oscillations of the suspended nanotube for node (red nanotube) and anti-node (blue nanotube). (b) Measured current oscillations under 6.9mW illumination at 633nm. (c) Zener tunneling model $I$-$V_{bias}$ curves obtained when the nanotube oscillates between 200K and 350K.

# Supplementary Document: Photo-thermal Self-oscillations in Cavity-Coupled Carbon Nanotube *pn*-Devices

*Moh. R. Amer*[*1,2], Tony Levi[3,5], and Stephen B. Cronin*[*3,4,5]

Department of Electrical Engineering[1]
University of California, Los Angeles, California, 90095
Center of Excellence for Green Nanotechnologies[2]
King Abdulaziz City for Science and Technology, Riyadh, 12612, Saudi Arabia
Departments of and Electrical Engineering[3], Material Science[4], Physics and Astronomy[5],
University of Southern California, Los Angeles, CA, 90089

**Table of Contents:**

## 1- Photothermal self-oscillations of a narrow band gap quasi-metallic carbon nanotube *pn* device:

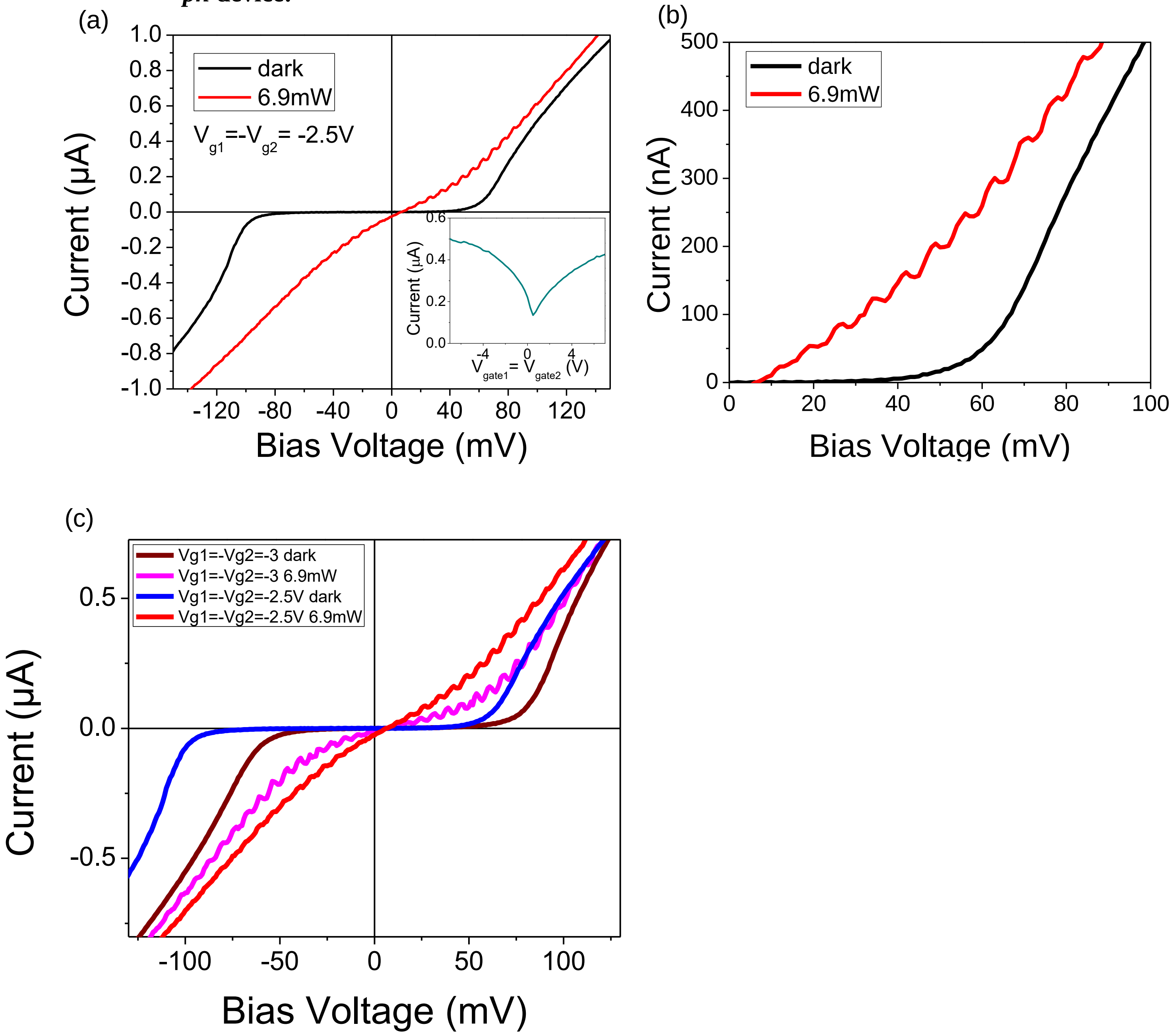

**Figure S1.** (a,b) $I$-$V_{bias}$ characteristics of a quasi-metallic carbon nanotube with and without optical excitatio. The inset shows the conductance of the nanotube as a function of gate voltage. (b) close-up of the data plotted in (a). (c) $I$-$V_{bias}$ characteristics of the suspended nanotube showing photo-thermal oscillations at different electrostatic doping.

## 2- Photothermal self-oscillations at room temperature for quasi-metallic carbon nanotube *pn* device.

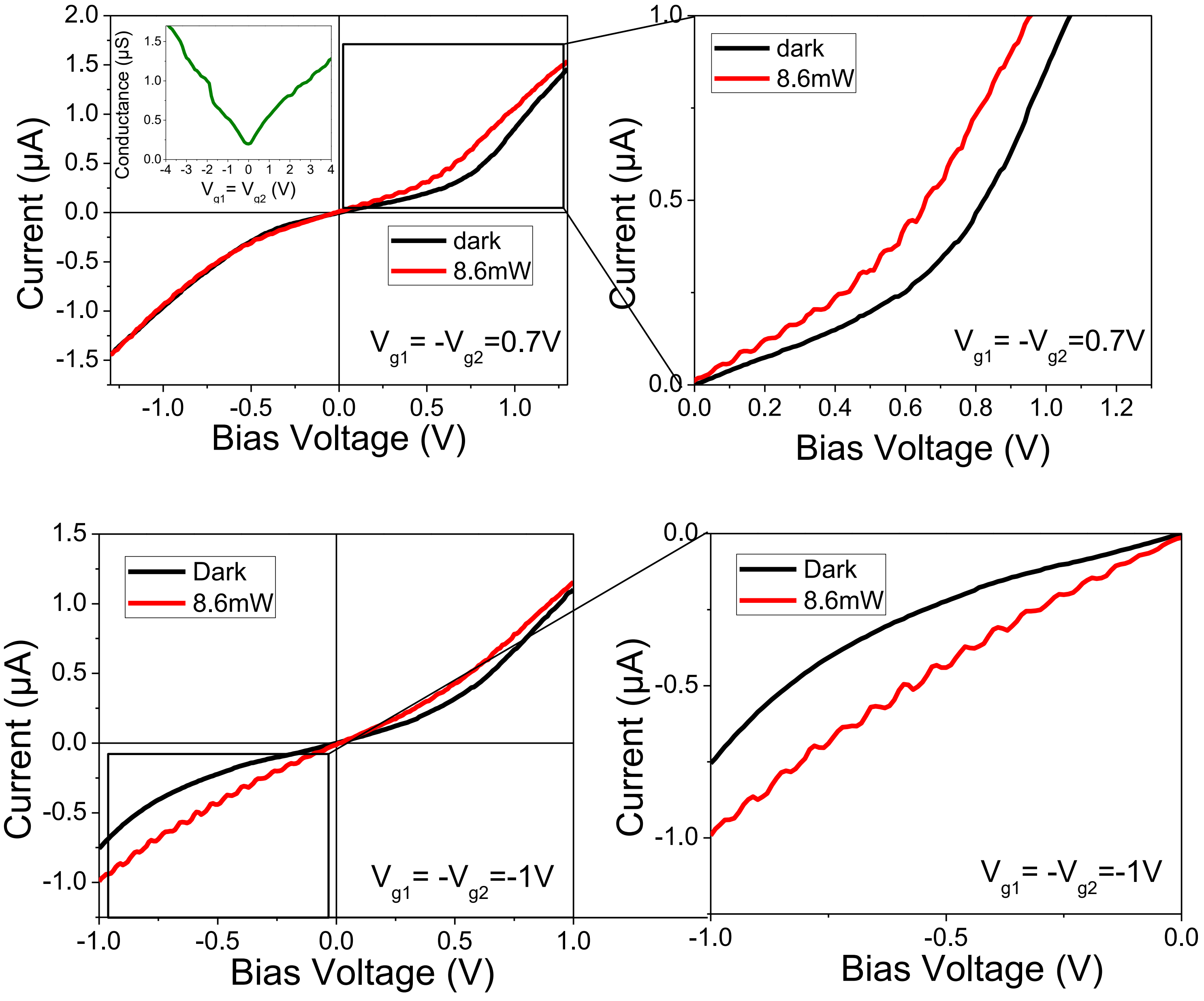

**Figure S2.** Photothermal oscillations at room temperature for a quasi-metallic nanotube *pn* device excited with 532nm laser for (a) $V_{g1}$=-$V_{g2}$=0.7 and (c) $V_{g1}$=-$V_{g2}$=-1V. The inset is the conductance curve of the nanotube. (b) and (d) close-up view of the oscillations in (a) and (c), respectively.

**a- Hysteresis in the forward and backward sweeps**

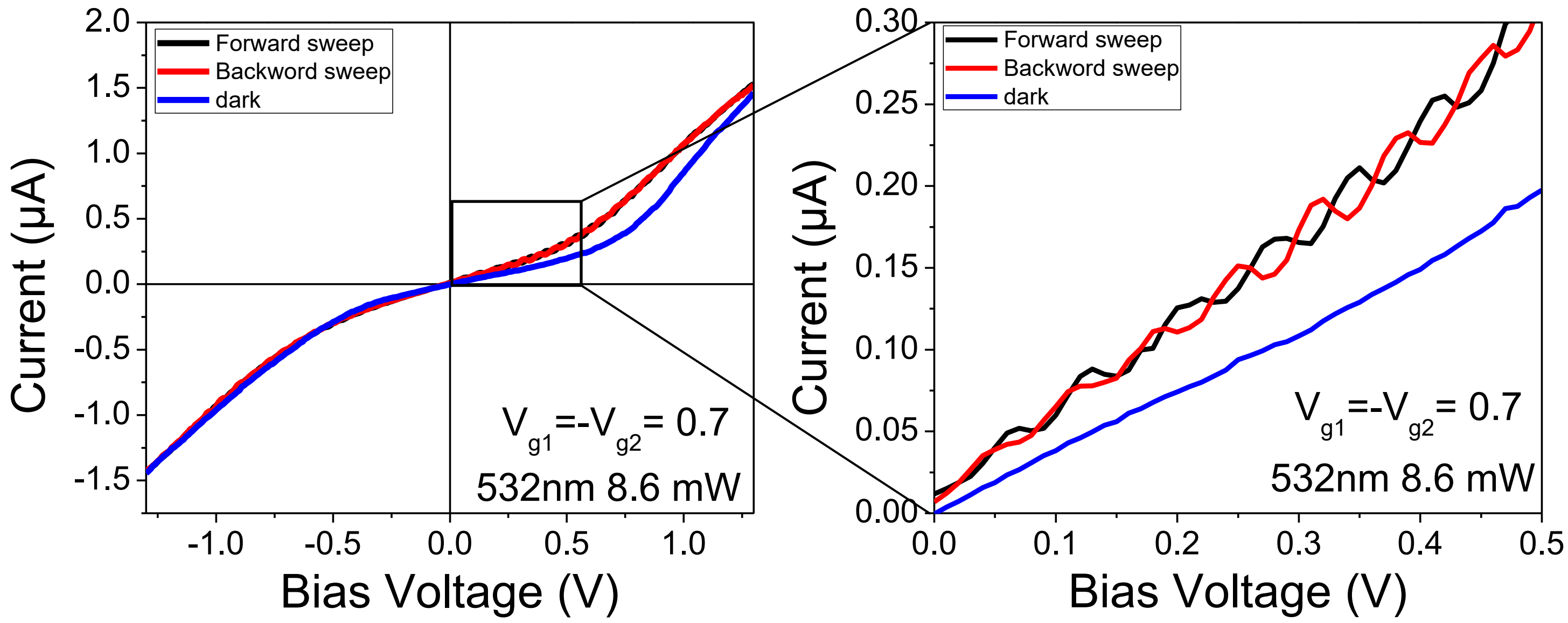

**Figure S3.** (a,b) *I*-$V_{bias}$ characteristics of the suspended nanotube at room temperature in the dark and when optically excited with 542nm laser. (b) Close-up view of the backward and forward sweeps showing hysteresis in the oscillations for the forward and backward sweeps, indicating photothermal oscillations.